\documentclass[fleqn,usenatbib,useAMS]{mnras}
\usepackage{newtxtext,newtxmath}
\usepackage[T1]{fontenc}
\usepackage{ae,aecompl}
\usepackage{graphicx}
\usepackage{amsmath}
\usepackage{amssymb}
\usepackage{multicol}
\usepackage{subfigure} 
\usepackage{bm}

\usepackage[T1]{fontenc}
\usepackage{ae,aecompl}
\usepackage{newtxtext,newtxmath}

\title[Machine learning to reveal gravitational lenses]{Machine learning and Kolmogorov analysis to reveal gravitational lenses}

\author[S. S. Mirzoyan et al.]{
S. S. Mirzoyan$^{1,2}$, H. Khachatryan$^2$, G. Yegorian$^2$ and V.G. Gurzadyan$^{2,3}$
\\
$^{1}$ Department of Physics, University of Zurich, Winterthurerstrasse 190, Zurich, Switzerland\\
$^{2}$ Center for Cosmology and Astrophysics, Alikhanian National Laboratory and Yerevan State University, Yerevan, Armenia\\
$^{3}$ SIA, Sapienza University of Rome, Rome, Italy
}

\date{Accepted 2019 August 6. Received 2019 July 31; in original form 2019 June 3}

\pubyear{2019}

\begin{document}
\label{firstpage}
\pagerange{\pageref{firstpage}--\pageref{lastpage}}
\maketitle

\begin{abstract}
We present an automated approach to detect and extract information from the astronomical datasets on the shapes of such objects as galaxies, star clusters and, especially, elongated ones such as the gravitational lenses. First, the Kolmogorov stochasticity parameter is used to retrieve the sub-regions that worth further attention. Then we turn to image processing and machine learning Principal Component Analysis algorithm to retrieve the sought objects and reveal the information on their morphologies. 
We show the capability of our automated method to identify distinct objects and to classify them based on the input parameters. A catalog of possible lensing objects is retrieved as an output of the software, then their inspection is performed for the candidates that survive the filters applied.
\end{abstract}


\section{Introduction}

The currently available various large scale sky surveys provide deep and well sampled astronomical datasets. In view of ever increasing amount of the digitized information and of the number vs their morphology of the involved sought objects, the development of efficient automatic processing of the datasets has no alternatives. Various automated methods have been described so far \cite{Al,Len,Sei} to process astronomical datasets. Among them machine learning and neural network techniques \cite{Hez,Hez1,Pet_Tor} are becoming ever more common and powerful application means in order, first, to separate the signals of astronomical object from noise and then to classify them according to certain criteria. 

The observational information on the gravitational lensing \cite{SEF,SJK,SKW} has become an important tool for tracing the large scale structure of the Universe, especially due to the ability for detection of the dark matter and even testing of modified gravity models, e.g. \cite{GSt,GSh}.  It enables to reveal the key properties of the extragalactic objects, both of the lensed ones and of those acting as lenses, as predicted already by Zwicky \cite{Zw}. Currently morphological variety of the lens caustics, starting from twin images of the quasar SBS 0957+561 up to Einstein rings, crosses, arcs, multiple images, are discovered and interpreted, see \cite{SKW,Tr} and references therein.    

The search of the gravitational lensing evidences in the galaxy surveys includes combined use of available observational information on the lensed images and the lens itself, i.e. the spectroscopy, photometry, color, morphology. The visual inspection, however, still remains among the important steps in the recognition of elongated/distorted structures as candidates for caustics with subsequent verification by other means \cite{Wis,Fr,SKW,LG,In,MMGM,Ni}. 
  
Below for the first time we apply automated strategy for detection of caustics of gravitational lensing using the method of Kolmogorov stochasticity parameter (KSP) \cite{K,UMN,MMS,FA}. That approach enables one to analyze signals which contain both correlated and random subsignals. Besides the study of generated sequences and modeling, that method has been already applied to the cosmic microwave background (CMB) datasets and enabled to separate e.g. the Galactic foreground, point sources (galaxies, quasars) from the cosmological signal, to analyze tiny properties of the latter \cite{GK_KSP,GK_void,GS,GA,cspot}. Among applications to datasets of quite different origin are e.g. the revealing of galaxy clusters in XMM-Newton's X-ray data \cite{Xray}, the effect of thermal trust perturbing the trajectories of laser ranging satellites \cite{GC}, the detection of somatic mutations in genomic sequences \cite{genom}.
 
In this methodical paper we give a description of a developed three-step approach, including its application to real astronomical data.  The aim of the paper is to show, first, the ability of filtering of the sub-regions that contain astronomical objects, then to identify and extract additional morphological information on those using the machine learning Principal Component Analysis algorithm.
   
\section{The method}
First, let us define the structure of the dataset under consideration. Consider an image given by 2D matrix of rectangular data, each $(x_{i}, y_{i})$ pair representing a pixel with intensity $I_{ij}$. 

For quantitative detection of gravitational arcs in astronomical datasets we apply 3 main steps:
\begin{enumerate}
\setcounter{enumi}{0}
\item Kolmogorov analysis,
\item object identification,
\item object classification.
\end{enumerate}

Note that, before the first step one should gain idea on data distribution creating the histogram of intensity. That procedure is also required for Kolmogorov maps which we obtain as result of KSP analysis \cite{K,UMN,MMS,FA}. This steps are needed to subtract the background from the original data, as well as to cut off the valuable data according certain criteria above given threshold.  The ``survived'' data are analysed for object finding and then for morphological parameter obtaining.

\section{Kolmogorov stochasticity parameter} \label{kolmogorov}

Consider a real-valued sequence of numbers $\{X_1,X_2,\dots, X_n\}$ represented in increasing order.  
One can define two distribution functions, i.e. an empirical distribution function as given \cite{K,UMN} 
\begin{eqnarray*}
F_n(x)=
\begin{cases}
0\ , & x<X_1\ ;\\
k/n\ , & X_k\le x<X_{k+1},\ \ k=1,2,\dots,n-1\ ;\\
1\ , & X_n\le x\ ,
\end{cases}
\end{eqnarray*}
and a theoretical (cumulative) distribution function as the probability  $F(x) = P\{X\le x\}.$
The difference of both distribution functions is represented by the Kolmogorov stochasticity parameter (KSP) $\lambda_n$ as $ \lambda_n=\sqrt{n}\ \sup_x|F_n(x)-F(x)|.$

Kolmogorov's theorem \cite{K} states that $\lim_{n\to\infty}P\{\lambda_n\le\lambda\}=\Phi(\lambda)\ $, with $\Phi(0)=0$, and where 
\begin{equation}
\Phi(\lambda)=\sum_{k=-\infty}^{+\infty}\ (-1)^k\ e^{-2k^2\lambda^2}\ ,\ \  \lambda>0\ ,\label{Phi}
\end{equation}
so that the function $\Phi$ is independent on the theoretical distribution function $F$. The form of Kolmogorov's function $\Phi$ determines $0.3\le\lambda_n\le 2.4$ for the KSP interval as the measure of degree of randomness of the above defined sequences \cite{UMN,MMS,FA}. The importance of this descriptor is that it is applicable even to sequences of few tens of length \cite{MMS}, which is not the case for most of statistical methods and is rather sensitive to the deviation from randomness. These features of the descriptor appear to be efficient at non-linear data analysis, see \cite{atto,R}.

We then applied KSP-analysis to the observational data of a strong lensed object, namely, to the SDP.81 galaxy \cite{Alma,Tam}. Our task is to find out whether KSP-method can distinguish the signal of a lensed object from that of the surrounding field, as it had enabled to separate the contribution e.g. of the Galactic disk from CMB in WMAP or Planck maps (see \cite{GA,GK_void,cspot}, also for further details of the application of KSP-method). 

We split the data field of size $\textit{M $\times$ N}$ into smaller sub-regions of, say $\textit{m $\times$ n}$ size, then we calculate KSPs for each sub-regions, composing so-called Kolmogorov map of the whole field. The knowledge on the original data distribution is crucial for two reasons: first, for choice of theoretical distribution function and, second, for background subtraction in the object identification step. 

\begin{figure}
\centering 
\includegraphics[width=0.4\textwidth]{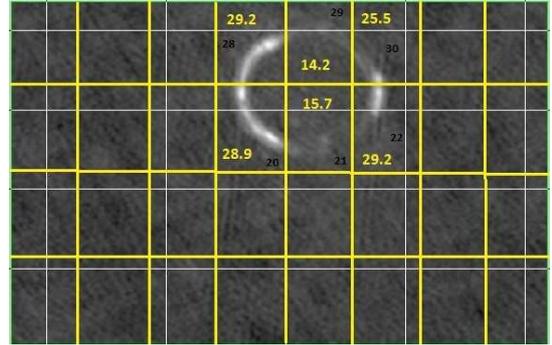}
\caption{The field of about [Dec=9'', RA=10''] of the galaxy SDP.81. The sub-regions of (84 x 110)-pixels with the
highest values of the KSP  are shown.}
\label{fig:sdp81}
\end{figure}

\begin{figure}
\centering 
\includegraphics[width=0.45\textwidth]{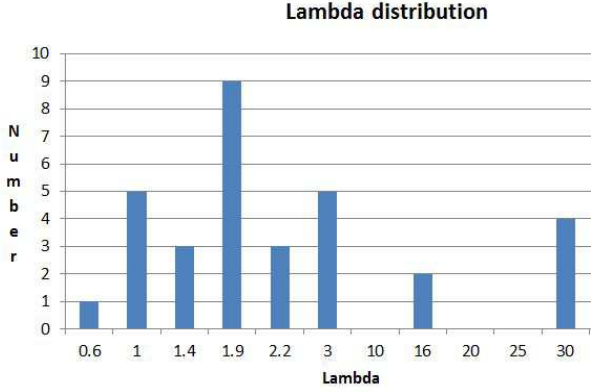}
\caption{The field of [Dec=9'', RA=10''] of the galaxy SDP.81. The sub-regions of (84 x 110)-pixels with the
highest values of the KSP  are shown.}
\label{fig:lambda_hist}
\end{figure}

The detailed analysis revealed the Gaussianity of the data, hence the theoretical cumulative function was taken as Gaussian (cf.\cite{GK_KSP}) and the results of the obtained values of KSP for each of the pixelized sub-regions are given in Figure $\ref{fig:sdp81}$. Before making conclusions it is worth to take a glance at the  $\lambda$ distribution of 32 sub-regions  (see Figure $\ref{fig:lambda_hist}$ ). The original data field of 672$\times$440 pixels was split into 32 sub-regions of 84$\times$110 pixels size. Therefore, in the Figure we show the histogram made of 32 $\lambda$ values, having mean value of around 1.9. One can clearly see that the majority of $\lambda$ values are between 0.5 and 2.2, which confirms the correctness of our assumption regarding the data Gaussianity. Higher values are due to the more regular structures in the sub-regions. The sub-regions that contain parts of lensing arcs have anomalously high  $\lambda$.

The further steps strongly depend on the data type: if the original image is sparse, it is appropriate to take the modal value (although, in this case the mean is close to modal) for $\lambda$ as background, otherwise, if the image is full of objects, the mean value for $\lambda$ might be considered. We filter KSP map with cut-off value defined as

\begin{equation}
\lambda_{thres} = \lambda_{0}+n*\sigma,
\end{equation}
where $\lambda_{0}$ is either mean or modal value of the entire field, $\sigma$ is the standard deviation, and finally $\textit{n}$ multiplier indicates how many $\sigma$-s we want to cut above the mean or modal value. This multiplier should rather be decided empirically.

Due to the data spread we get large value of standard deviation, therefore $n=1$ is a good choice for multiplier. This results in 6 sub-regions to pass our filter. By next step we are going to identify objects in those regions.

\section{Object identification}

This section we devote to description of the object finding algorithm. Let us first define the object. It is a set of one-connected pixels that are isolated from other sets of connected pixels. And as far as we are looking for astronomical object composing pixels, those pixels should have somewhat higher values of intensity. Therefore using our knowledge about original data (distribution, mean or modal value and standard deviation) we apply cut-off technique and maintain only pixels that survive the filter (survived data).
Having a list of those pixels  p($x_{i}, y_{i}$) we define Moore neighborhood $\textit{N(p($x_{i}, y_{i}$))}$ for each of them, that is the set of all pixels that are orthogonally or diagonally-adjacent to the region of interest and the region of interest itself may or may not be considered part of the Moore neighborhood \cite{Moore}. In other words, for the given pixel with ($x_{i}, y_{i}$) coordinates, the neighbors are those with coordinates obeying the following inequality

\begin{equation}
\sqrt{(x_{i}-x_{j})^2+(y_{i}-y_{j})^2} \leq \sqrt{2}.
\end{equation}

There are different approaches to check the path connectivity between two pixels \cite{Soi,Len}. However we propose the simple algorithm that we came up with.

First of all we construct the lists of neighbors for each pixel. As far as we search those pixels from the survived data, the boundary pixels of objects do not contain the full set of neighbors.  After this step, we take an auxiliary list $\textbf{A}$ that, in the end is supposed to contain the sets of pixels for different objects (it is set of those sets), and compare its content with every $\textbf{N(p)}$. 

$a) \,$ If $A \cap N(p) =\{p_{k} ... p_{m}\}$ ,  this means that current pixel belongs to object $\to$ add $A \setminus N(p)$ into A;

$b) \,$ Otherwise, if $A \cap N(p) =\emptyset$,  $\to$ iterate over elements of A, if finds intersecting lists, $\to$ go to $a) \,$, completes, then breaks, otherwise if it iterates till the end, just adds a new set into A with the current N(p) elements and then breaks.

This procedure results with the set of isolated objects for each sub-regions. 

\section{Object classification}

Object classification is the final and the most important step of this algorithm. Up to now we described our approach to the automated arc searching using the KSP-method that helps saving computational efforts by filtering the fields with valuable information. Then, we developed an object finding algorithm and now we are going to give detailed description of the object classification algorithm.

First of all, let us recall that, the gravitational lensing arcs are elongated objects - particularly arcs of a circle originated due to the lens distribution along the line-of-sight. Hence, our task is to search for that kind of objects, sometimes as tiny as PSF. Some galactic spiral arms are similar to arcs, so can be mutually confused, if not the typically significant thickness of arms. 

Let the isolated object be composed of set of pixels

\begin{equation}
p_{i} = x_{i},y_{i} : i \in N.
\end{equation}

The mathematical moment of $\textit{m+n}$ order  is defined as 

\begin{equation}
\sum_{i=0}^{M}\sum_{j=0}^{N} x_{i}^{m} y_{j}^{n} f(x,y).
\end{equation}

In order to calculate the object's total intensity and the center we calculate 0-th and 1-th order moments, correspondingly,

\begin{equation}
sum_{x} = \sum_{i=0}^{M} x_{i} f(x,y)
\end{equation}

and 

\begin{equation}
sum_{y} = \sum_{j=0}^{N} y_{i} f(x,y)
\end{equation}

and then, we get the center of the object as follows

\begin{equation}
c_{x} = \frac{sum_{x}}{\mu_{0}},
\end{equation}

\begin{equation}
c_{y} = \frac{sum_{y}}{\mu_{0}},
\end{equation}
where $\mu_{0}$ is the total intensity of the object.
The second order moment is defined as  $2 \times 2$ matrix with components
\begin{equation}
Cov_{jk} = \frac{1}{m} \sum f(x_{i},y_{i}) (x_{i}^j-c_{x})(y_{i}^k-c_{y}).
\end{equation}

This matrix is called $\textit{covariance}$ matrix. It is a symmetric matrix having variance values for $\textit{x}$ and $\textit{y}$ independent variables in its diagonal. While the variance is defined for a set of variables and refers to the spread of data around its mean, the covariance refers to the measure of the directional relationship between two random variables. Hence it can be used to investigate the properties of isolated objects, whether elongated or more regular. To this end we calculate the  $\textit{eigenvalues}$ of matrix $Cov_{jk}$ and  $\textit{eigenvectors}$. Whereas the eigenvectors define the principal components of the data, the eigenvalues are the scales along those principal components.

A $2 \times 2 $ matrix can have at most two eigenvalues, say $\lambda_{1}$ and $\lambda_{2}$, and if $\lambda_{1}$ $>$ $\lambda_{2}$, then
\begin{equation}
e = \frac{\lambda_{1}-\lambda_{2}}{\lambda_{1}}
\end{equation}
is the eccentricity of the object. Our software includes threshold  parameter for eccentricity and thickness of the elongated object. So, depending on data we study, those threshold values vary and might be induced empirically. Hence, at the end, our software outputs a list of objects with a diagnosis about object type. Of course the results depend on our choice of $e_{thresh}$ and $d_{thick}$. The objects for which 
\begin{equation}
e \geq e_{thresh}
\end{equation}
and 
\begin{equation}
\lambda_{2} \leq d_{thick}
\end{equation}
only appear in the final list.

\section{Results}

Below we introduce a table that contains a list of isolated objects. We set the threshold value for eccentricity 0.35, therefore in the table we have objects with small eccentricities. The reason we keep objects with small eccentricities is that, in the first step we split the whole field into smaller sub-regions, and as it has been shown in Table $\ref{table:obj}$, the lensing structure is separated between parts. This assumes an extra work of grouping the parts of the same structure together, but at the same time we make profit when with Kolmogorov analysis we avoid performing the search of structures in the entire field. Indeed, we just look for objects in the fields that survive $\lambda$ filtering. 

\begin{table}[htb]
\centering
\caption{List of objects: coordinates of centers, eccentricity and field numbers. 2$\sigma$  is used for data cut-off.}
\label{table:obj}
 \begin{tabular}{||c | c | c | c||} 
 \hline
 X center & Y center & Eccentricity & Field Number \\ [0.5ex] 
 \hline\hline
 306.00 & 289.63 & 0.9639 & 20 \\ 
 \hline
 327.22 & 259.62 & 0.666 & 20\\
 \hline
 308.22 & 271.63 & 0.666 & 20 \\
 \hline
 344.52 & 256.28 & 0.8309 & 21 \\
 \hline
 348.22 & 247.62 & 0.666 & 21 \\
 \hline
 338.75 & 252.67 & 0.6855 & 21 \\ 
 \hline
 391.87 & 251.87 & 0.375 & 21\\
 \hline
 392.38 & 263.70 & 0.5264 & 21 \\
 \hline
 458.49 & 315.27 & 0.8417 & 22 \\
 \hline
 309.15 & 370.75 & 0.9433 & 28 \\
 \hline
 343.40 & 393.57 & 0.7568 & 29 \\ 
 \hline
 415.66 & 387.00 & 0.62 & 29\\
 \hline
 412.51 & 399.09 & 0.672 & 29 \\
 \hline
 390.16 & 403.96 & 0.3564 & 29 \\
 \hline
 458.41 & 340.05 & 0.542 & 30 \\ [1ex] 
 \hline
\end{tabular}
\end{table}

\begin{figure}
\centering 
\includegraphics[width=0.45\textwidth]{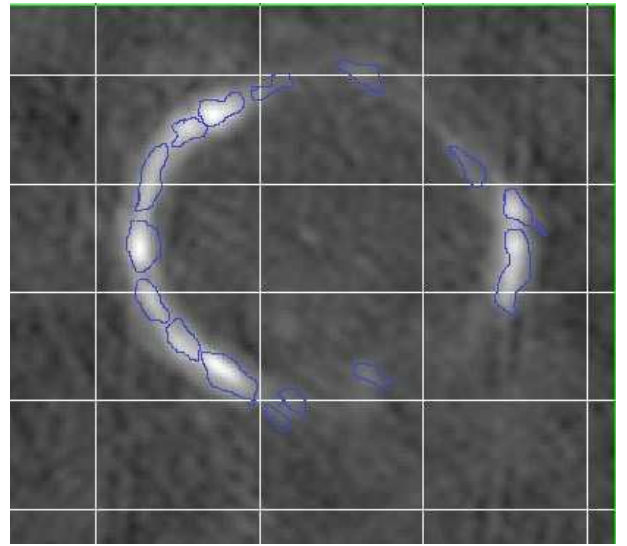}
\caption{The identified objects are shown. Blue contours show approximate structures of objects above 3-$\sigma$, some of the objects are combined together.}
\label{fig:ring}
\end{figure}

\begin{figure} 
\subfigure[\label{fig:1s} {\small Objects above 1-$\sigma$}.]{ 
\includegraphics[width=0.45\linewidth]{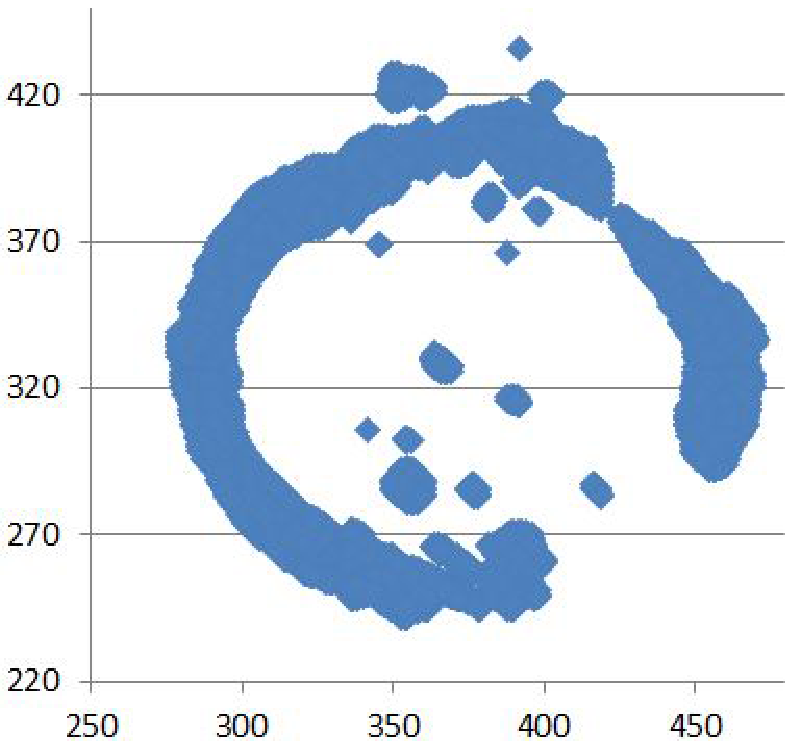}}
\hfill 
\subfigure[\label{fig:15s} {\small Objects above 1.5-$\sigma$}]{ 
\includegraphics[width=0.45\linewidth]{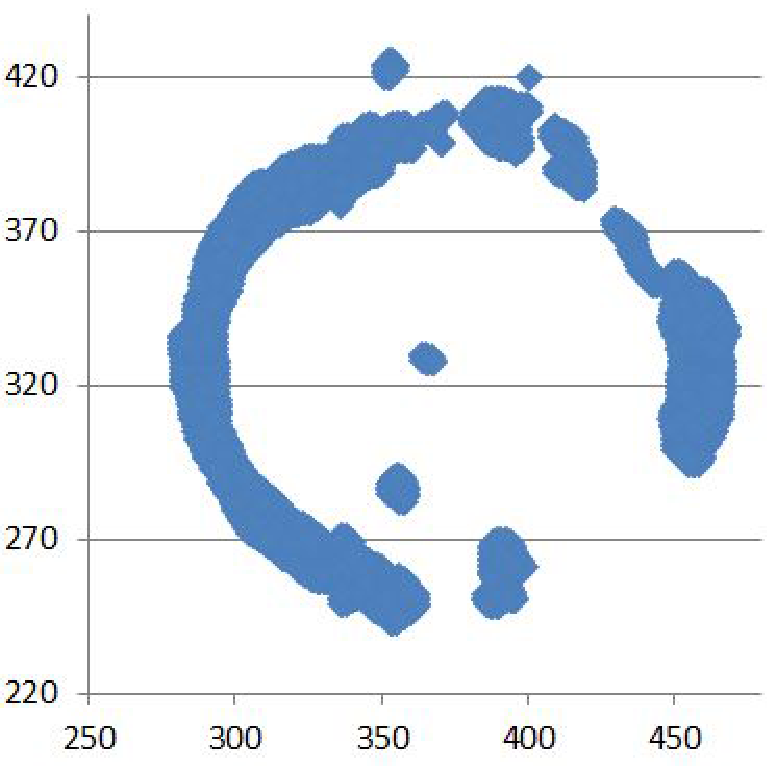}}
\vfill
\subfigure[\label{fig:2s} {\small Objects above 2-$\sigma$}]{ 
\includegraphics[width=0.45\linewidth]{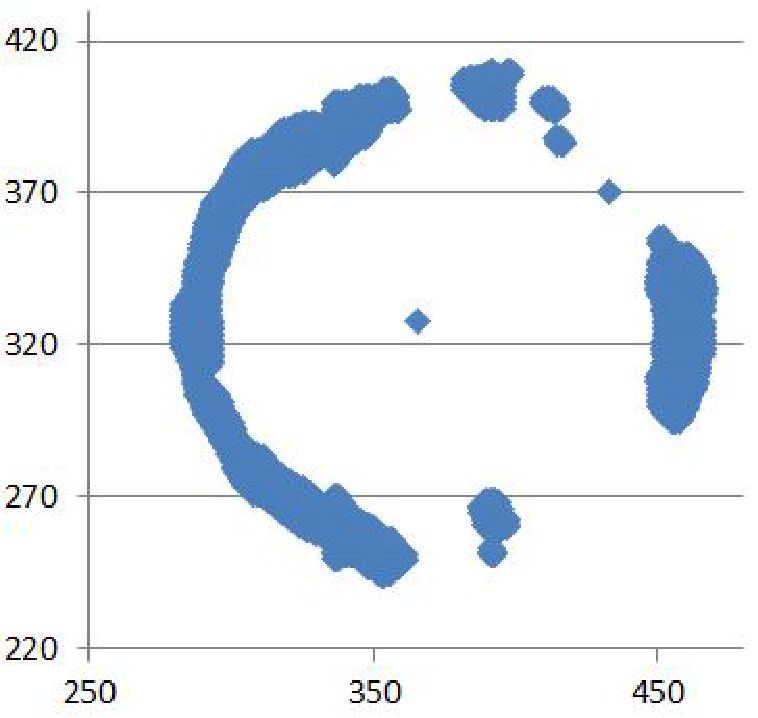}}
\hfill 
\subfigure[\label{fig:3s} {\small Objects above 3-$\sigma$}]{ 
\includegraphics[width=0.45\linewidth]{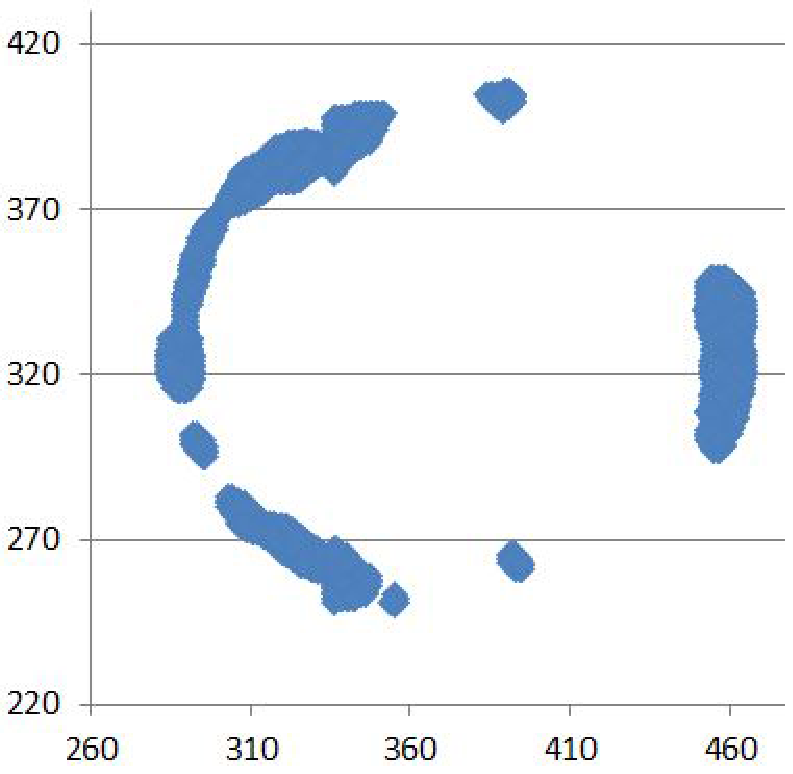}}
\caption{Objects identified for different intensity thresholds.
         Reconstruction of lensing event by identified objects. 
         Each image corresponds to different intensity thresholds 
         (1-$\sigma$, 1.5-$\sigma$, 2-$\sigma$, 3-$\sigma$.)}
\label{fig:alls}
\end{figure}

From the Table $\ref{table:obj}$ one can see that the first 3 objects belong to the same field, and from the first glance those are in the same big arc. However attentive inspection of Fig. $\ref{fig:ring}$, in particular of sub-region 20, makes clear that along the big arc there is  variation of pixel intensities, so that some part does not pass the filter and the visibly continuous arc splits to several disjoint arcs of smaller size. In Fig. $\ref{fig:alls}$ we represent the results of our software for different cut-off values of intensities, particularly, Fig. $\ref{fig:1s}$  corresponds to 1-$\sigma$, Fig. $\ref{fig:15s}$ to 1.5-$\sigma$,  Fig. $\ref{fig:2s}$ to 2-$\sigma$ and Fig. $\ref{fig:3s}$ to 3-$\sigma$.

Note that, the results of this method depend on the quality of image. Due to the sensitivity of KSP-method,  even the hidden from eye signals can be revealed.  Even if the background of field is quasi-uniform at high level, small perturbations can be detected by KSP-method. An efficient  Object identification and Object classification are matter of careful understanding of the data available. Indeed, for those steps it is crucial to know the data distribution, as well as its mean or modal values, and to perform a proper cut-off to avoid loosing valuable information. 
\begin{figure*}
\includegraphics[width=7in]{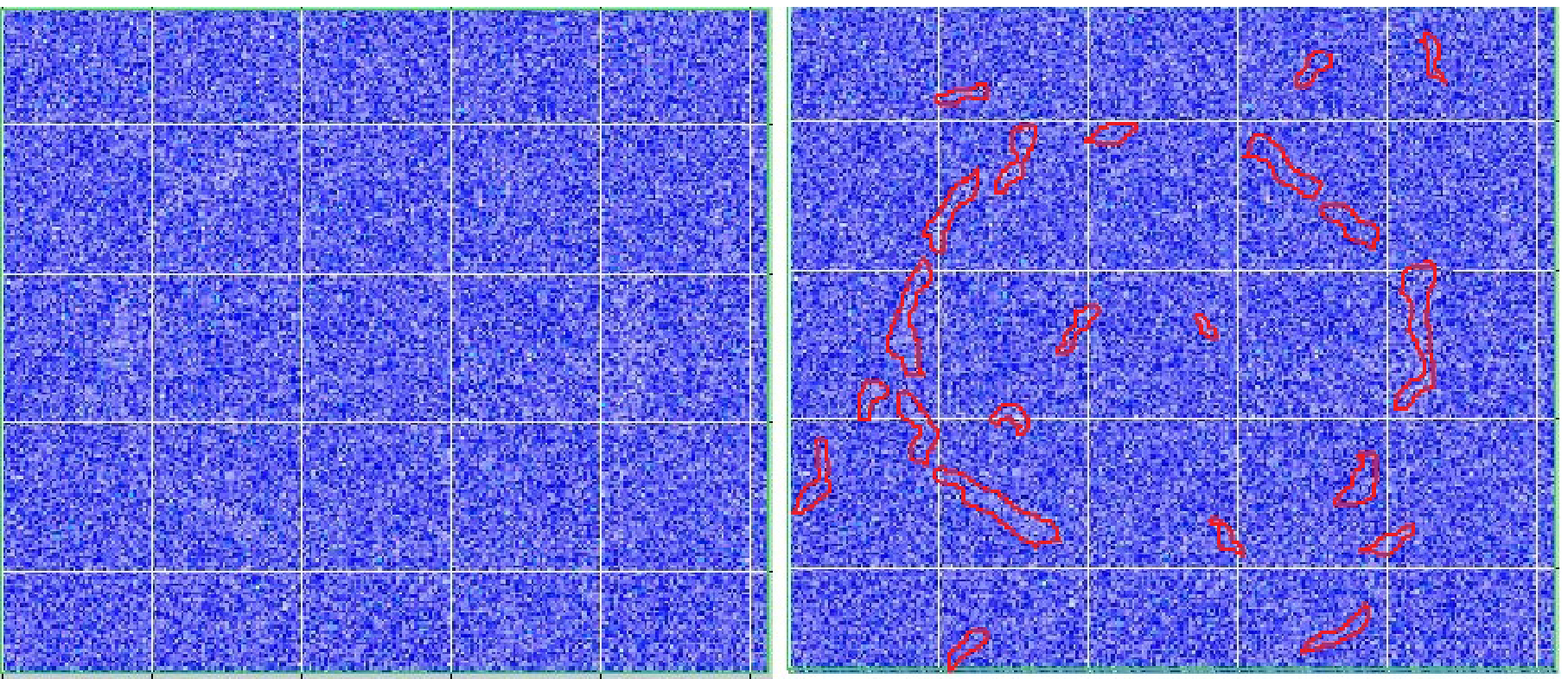}
\caption{Simulated image of lensing with 0.5-$\sigma$ above background level (left plot). The right plot shows the results of the search algorithm. Except clearly exhibited lensing arcs the method identifies other elongated objects, which should be investigated separately. Some of most emphasized elongated objects are shown in the right plot.}
\label{fig:simul}
\end{figure*}

Additionally, the ability of the suggested method to reveal "low visibility" structures can be seen from the following example.  We simulated images of different statistical significance. Namely, even if the object pixels' intensities are only 0.5-$\sigma$ above the background level, this method is still able to identify elongated objects (Fig. $\ref{fig:simul}$). Indeed, it is hardly possible to notice structures in the initial simulated image (left plot), whereas the right image presents the results retrieved  by the algorithm described above. Of course, as one might expect, other structures may be retrieved as well as shown in the right plot and their possible association to the lensing structure candidates should be investigated additionally.

\section{Conclusions}

We advanced an automated method of search for isolated objects in astronomical datasets, that is, extraction of valuable information based on statistical properties of data and Moore's neighborhood algorithm.

First, the split of pixelized regions containing the gravitational lens images showed statistically notable difference regarding the Kolmogorov function with respect to their average value in the surrounding sky. 
Then, we extracted morphological information on the extracted objects applying Principal Component Analysis strategy. This enabled us to classify objects among elongated or regular structures. The efficiency of the method is illustrated via simulation of low significance structures.  

Observational surveys offer huge amount of data and development of any new tool for automated search for given category of objects can be important. Particularly, the Kolmogorov analysis in addition to the problems mentioned in the Introduction can be applied to test the isotropy of sky distribution of gamma-ray bursts. That problem has been studied by various methods, not always with identical conclusions (see \cite{RC,And,Rip} and references therein), and is of remarkable cosmological importance. Image processing and machine learning algorithms are currently becoming conventional tools for astronomical datasets, to extract ever more refined information. In this paper we considered an example of combination of statistical methods and machine learning algorithms to reveal certain structures in astronomical datasets.

\label{lastpage}
\section{Acknowledgments}

We acknowledge the use of data from http://almascience.nrao.edu/aq/.


\begin{thebibliography}{99}

\bibitem[\protect\citeauthoryear{Alard}{2006}]{Al} Alard C., 2006, arXiv:astro-ph/0606757
\bibitem[\protect\citeauthoryear{ALMA }{2015}]{Alma} ALMA Partnership, Vlahakis, C. et al, 2015, ApJL, 808, L4
\bibitem[\protect\citeauthoryear{Andrade et al}{2019}]{And} Andrade U., Bengaly C.A.P., Alcaniz J.S., Capozziello S., 2019,  arXiv:1905.08864
\bibitem[\protect\citeauthoryear{Arnold}{2008}]{UMN} Arnold V.I., 2008, Uspekhi Mat.Nauk, 63, 5  
\bibitem[\protect\citeauthoryear{Arnold}{ 2009a}]{MMS} Arnold V.I., 2009a, Trans. Moscow Math. Soc., 70, 31 
\bibitem[\protect\citeauthoryear{Arnold}{ 2009b}]{FA} Arnold V.I., 2009b, Funct. An. Other Math. 2, 139
\bibitem[\protect\citeauthoryear{Atto et al}{ 2013}]{atto} Atto A.M., Berthoumieu Y., Megret R., 2013, Entropy, 15, 4782
\bibitem[\protect\citeauthoryear{Frey et al}{ 2003}]{Fr} Frey, S., Mosoni, L., Paragi, Z., and Gurvits, L. I. 2003, MNRAS, 343, L20 
\bibitem[\protect\citeauthoryear{Gurzadyan \& Kocharyan}{ 2008}]{GK_KSP} Gurzadyan V.G. \& Kocharyan A.A., 2008, A\&A 492, L33  
\bibitem[\protect\citeauthoryear{Gurzadyan \& Kocharyan}{ 2009}]{GK_void} Gurzadyan V.G. \& Kocharyan A.A., 2009, A\&A, 493, L61 
\bibitem[\protect\citeauthoryear{Gurzadyan \& Stepanian}{ 2018}] {GSt} Gurzadyan V.G., Stepanian A., 2018, Eur. Phys. J. C, 78,  869
\bibitem[\protect\citeauthoryear{Gurzadyan \& Stepanian}{ 2019}] {GSh} Gurzadyan V.G., Stepanian A., 2019, Eur. Phys. J. C, 79,  568 
\bibitem[\protect\citeauthoryear{Gurzadyan et al}{ 2008}]{GS} Gurzadyan V.G., Starobinsky A.A., et al, 2008, A\&A, 490, 929 
\bibitem[\protect\citeauthoryear{Gurzadyan et al}{ 2009}]{GA} Gurzadyan V.G., Allahverdyan A.E. et al, 2009, A\&A, 497, 343
\bibitem[\protect\citeauthoryear{Gurzadyan et al}{ 2011}]{Xray} Gurzadyan V.G., Durret F. et al, 2011, Europhys. Lett.  95, 69001
\bibitem[\protect\citeauthoryear{Gurzadyan et al}{ 2013}]{GC} Gurzadyan V.G., Ciufolini I. et al, 2013, Europhys. Lett., 102, 60002 
\bibitem[\protect\citeauthoryear{Gurzadyan et al}{ 2014}]{cspot} Gurzadyan V.G., Kashin A.L. et al, 2014, A\&A, 566, A135
\bibitem[\protect\citeauthoryear{Gurzadyan et al}{ 2015}]{genom} Gurzadyan V.G., Yan H. et al, 2015, Roy. Soc. Open Science, 2, 150143
\bibitem[\protect\citeauthoryear{Hezaveh et al}{ 2016}]{Hez} Hezaveh, Y. D., Dalal, N. et al, 2016, ApJ, 823, id. 37
\bibitem[\protect\citeauthoryear{Hezaveh et al}{ 2017}]{Hez1} Hezaveh, Y. D. et al, 2017, Nature, 548, 555
\bibitem[\protect\citeauthoryear{Inoue et al}{ 2015}]{In}Inoue, K. T., Minezaki, T., Matsushita, S., and Chiba, M. 2015, arXiv:1510.00150
\bibitem[\protect\citeauthoryear{Kolmogorov}{ 1933}]{K} Kolmogorov A.N., 1933, G.Ist.Ital.Attuari, 4, 83
\bibitem[\protect\citeauthoryear{Lenzen et al}{ 2004}]{Len} 	Lenzen, F., Schindler, S., Scherzer, O., 2004, A\&A, 416, 391
\bibitem[\protect\citeauthoryear{Lopez-Caniego et al}{ 2013}]{LG}Lopez-Caniego, M., Gonzalez-Nuevo, J. et al, 2013, MNRAS, 430, 1566
\bibitem[\protect\citeauthoryear{Mediavilla et al}{ 2016}]{MMGM} Mediavilla E., Munoz J., Garzon F., Mahoney T.J., (Eds.),  Astrophysical Applications of Gravitational Lensing, Cambridge University Press, Cambridge (2016)
\bibitem[\protect\citeauthoryear{Moore}{ 1964}]{Moore} Moore E.F.,  Sequential Machines, Selected Papers (1964)
\bibitem[\protect\citeauthoryear{Nierenberg et al}{ 2017}]{Ni}Nierenberg, A.M., Treu, T. et al, 2017, arXiv:1701.05188
\bibitem[\protect\citeauthoryear{Petrillo et al}{ 2017}]{Pet_Tor} Petrillo, C. E., Tortora, C., 2017, MNRAS, 472, 1129
\bibitem[\protect\citeauthoryear{Ripa \& Shafieloo}{ 2019}]{Rip} Ripa J., Shafieloo A., 2019, MNRAS, 486, 3027 
\bibitem[\protect\citeauthoryear{Rossmanith}{ 2013}]{R} Rossmanith G., Non-linear Data Analysis on the Sphere, Springer (2013) 
\bibitem[\protect\citeauthoryear{Ruggeri \& Capozziello}{ 2016}]{RC} Ruggeri A.C., Capozziello S., 2016, ApSS, 361, 279
\bibitem[\protect\citeauthoryear{Schneider et al}{ 1992}]{SEF} Schneider, P., Ehlers, J., Falco, E. E., Gravitational Lenses. Springer, Berlin (1992)
\bibitem[\protect\citeauthoryear{Schneider et al}{ 2006}]{SKW} Schneider P. , Kochanek C.,  Wambsganss J., 2006, Gravitational Lensing: Strong, Weak and Micro: Saas-Fee Advanced Course 33, Springer, Berlin (2006)
\bibitem[\protect\citeauthoryear{Seidel \& Bartelmann}{ 2007}]{Sei} Seidel, G., Bartelmann, M., 2007, A\&A, 472, 341
\bibitem[\protect\citeauthoryear{Soille}{ 2003}]{Soi} Soille, P.,  Morphological Image Analysis,  (Springer) (2003)
\bibitem[\protect\citeauthoryear{Straumann et al}{ 1998}]{SJK} Straumann, N.,  Jetzer, Ph.,  Kaplan, J., Topics on gravitational Lensing, Napoli series on Physics and Astrophysics, 1, Naples (1998)  
\bibitem[\protect\citeauthoryear{Tamura}{ 2015}]{Tam} Tamura Y., Oguri M. et al, 2015, PASJ, 67, id.727
\bibitem[\protect\citeauthoryear{Treu}{ 2010}]{Tr} Treu, T., 2010, ARAA, 48, 87
\bibitem[\protect\citeauthoryear{Wisotzki et al}{ 2002}]{Wis} Wisotzki, L., Schechter, P. L. et al, 2002, A\&A, 395, 17 	
\bibitem[\protect\citeauthoryear{Zwicky}{ 1937}]{Zw} Zwicky F. 1937, Phys. Rev. 51, 290

\end{thebibliography}
\end{document}